\documentclass[]{spie}

\usepackage{amsmath,amsfonts,amssymb}
\usepackage{graphicx}
\usepackage{setspace}
\usepackage{tocloft}
\usepackage[colorlinks=true, allcolors=blue]{hyperref}

\title{SHARP - A near-IR multi-mode spectrograph conceived for MORFEO@ELT}

\author[a]{P. Saracco}
\author[a]{P. Conconi}
\author[b]{C. Arcidiacono}
\author[c]{E. Portaluri}
\author[a,d]{H. Mahmoodzadeh}
\author[e]{V. D'Orazi}
\author[f]{D. Fedele}
\author[g]{A. Gargiulo}
\author[h]{E. Vanzella}
\author[g]{P. Franzetti}
\author[a]{I. Arosio}
\author[a]{L. Barbalini}
\author[a]{G. Lops}
\author[a]{E. Molinari}
\author[i]{E. Cascone}
\author[i]{V. Cianniello}
\author[i]{D. D'Auria}
\author[i]{V. De Caprio}
\author[c]{I. Di Antonio}
\author[c]{B. Di Francesco}
\author[c]{G. Di Rico}
\author[i]{C. Eredia}
\author[g]{M. Fumana}
\author[b]{D. Greggio}
\author[h]{G. Rodeghiero}
\author[a]{M. Scalera}
\author[i]{J. M. Alcal\'a}
\author[g]{S. Bisogni}
\author[l]{R. Bonito}
\author[e]{G. Bono}
\author[i]{A. Caratti o Garatti}
\author[m,b,r]{E. Dalla Bont\`a}
\author[i]{M. Dall'Ora}
\author[n]{G. Fiorentino}
\author[f]{A. R. Gallazzi}
\author[l]{M. Guarcello}
\author[i]{L. Izzo}
\author[i]{F. La Barbera}
\author[o]{C. Lardo}
\author[a]{M. Longhetti}
\author[p]{A. Longobardo}
\author[f]{L. Magrini}
\author[g]{C. Mancini}
\author[p]{A. Mura}
\author[n]{E. Piconcelli}
\author[m]{A. Pizzella}
\author[f]{L. Podio}
\author[g]{M. Polletta}
\author[l]{L. Prisinzano}
\author[q,n]{F. Ricci}
\author[i]{V. Ripepi}
\author[f]{V. Roccatagliata}
\author[g]{G. Vietri}
\affil[a]{INAF - Osservatorio Astronomico di Brera, Via Brera 28, Milano, Italy, 20121}
\affil[b]{INAF - Osservatorio Astronomico di Padova, Vicolo Osservatorio 5, Padova, Italy, 35122}
\affil[c]{INAF - Osservatorio Astronomico d'Abruzzo,Via Mentore Maggini, Teramo, Italy, 64100}
\affil[d]{Politecnico di Milano, Piazza Leonardo da Vinci 32, Milano, Italy, 20133}
\affil[e]{Universita' degli Studi di Roma Tor Vergata, Via Cracovia 50, Roma, Italy, 00133}
\affil[f]{INAF - Osservatorio Astrofisico di Arcetri, Largo E. Fermi 5, Firenze, Italy, 50125}
\affil[g]{INAF - IASF, Via Alfonso Corti 12, Milano, Italy, 20133}
\affil[h]{INAF - Osservatorio di Astrofisica e Scienza dello Spazio, via Gobetti 93/3, Bologna, Italy, 40129}
\affil[i]{INAF - Osservatorio Astronomico di Capodimonte, Salita Moiariello 16, Napoli, Italy, 80131}
\affil[l]{INAF - Osservatorio Astronomico di Palermo, Piazza del Parlamento 1, Palermo, Italy, 90134}
\affil[m]{Universita' degli Studi di Padova, Via F. Marzolo 8, Padova, Italy, 35131}
\affil[n]{INAF - Osservatorio Astronomico di Roma, Via Frascati 33, Monte Porzio Catone (RM), Italy, 00078}
\affil[o]{Universita' degli Studi di Bologna, via Piero Gobetti 93/2, Bologna, Italy, 40129}
\affil[p]{INAF - IAPS, Via del Fosso del Cavaliere 100, Roma, Italy, 00133}
\affil[q]{Dipartimento di Matematica e Fisica, Universit\'a Roma Tre, Via della Vasca Navale 84, Roma, Italy, 00146}
\affil[r]{Jeremiah Horrocks Institute, University of Central Lancashire, Preston, PR1 2HE, UK}

\authorinfo{Send correspondence to Paolo Saracco: paolo.saracco@inaf.it}

\begin{document} 
\maketitle

\begin{abstract}
The Extremely Large Telescopes (ELTs), thanks to their large apertures and cutting-edge
 Multi-Conjugate Adaptive Optics (MCAO) systems, promise to deliver sharper and deeper 
 data even than the JWST. 
SHARP is a concept study for a near-IR (0.95-2.45 $\mu$m) spectrograph conceived to fully exploit the collecting area and the angular resolution of the upcoming generation of ELTs. In particular, SHARP is designed for the 2nd port of MORFEO@ELT.
Composed of a Multi-Object Spectrograph, NEXUS, and a multi-Integral Field Unit, VESPER,
MORFEO-SHARP will deliver high angular ($\sim$30 mas) and spectral (R$\simeq$300, 2000, 6000, 17000) resolution, outperforming NIRSpec@JWST (100 mas).
SHARP will enable studies of the nearby Universe and the early Universe in unprecedented detail.
NEXUS is fed by a configurable slit system deploying up to 30 slits with $\sim$2.4" length and
adjustable width, over a field of about 1.2’$\times$1.2’ (35 mas/pix). 
Each slit is fed by an inversion prism able to rotate by an arbitrary angle the field
 that can be seen by the slit.
VESPER is composed of 12 probes of 1.7''$\times$1.5'' each (spaxel 31 mas)
probing a field 24”$\times$70”. 
SHARP is conceived to exploit the ELT aperture reaching the faintest flux and the sharpest angular 
resolution by joining the sensitivity of NEXUS and the high spatial sampling of VESPER to MORFEO
capabilities. 
This article provides an overview of the scientific design drivers, their solutions, and the resulting 
optical design of the instrument achieving the required optical performance.
\end{abstract}



\begin{spacing}{1}   

\section{Introduction}
\label{sect:intro}  
\subsection{Overview}
The Extremely Large Telescopes (ELTs) are expected to
provide unique and unparalleled data, deeper than that of the JWST and
from four (GMT telescope) to six (ESO-ELT) times sharper. 
These capabilities result from the synergy between the
large apertures of these telecopes and the Multi-Conjugate
Adaptive Optics (MCAO) systems.
In Particular, the ESO's ELT will provide unique data thanks to the world's 
largest aperture and the state-of-the-art MCAO system MORFEO\cite{ciliegi22,ciliegi21},
designed to uniformly
correct atmospheric turbulence over an encircled
area of diameter 160". 
This capability enables supported instruments to delve into the Universe with 
unparalleled depth and clarity. Initially working with the ELT's first-light instrument,
MICADO\cite{davies21,davies18}, a near-infrared camera, MORFEO is also engineered to feed
 a second instrument in the future. A spectrograph intended for
the 2nd port of the MCAO system MORFEO@ELT will be the most
powerful instrument of the JWST\cite{gardner23} era, able to go beyond NIRSpec \cite{jakobsen22} at JWST.

SHARP\footnote{http://sharp.brera.inaf.it} is the concept study of a multi-mode Near-IR spectrograph designed for 
the 2nd port of MORFEO@ELT.
It is intended to be submitted in ESO's next call for new instrumentation. 
Coupled with MORFEO, SHARP delivers unprecedented high angular 
($\sim$30 mas) and spectral resolution (R=300, 2000, 6000, 17000), outperforming NIRSpec@JWST
(100 mas). 
SHARP aims at being the near-IR nearly diffraction limited multi-object 
spectrograph of ELT.

\subsection{Why SHARP: scientific drivers and resulting requirements}
Understanding and reconstructing how baryonic matter assembled at early cosmic times to form 
the first stars, galaxies and structures, and how these evolved over cosmic time 
are the main scientific drivers behind the development of SHARP. 
Some of the fundamental issues that determined SHARP's main requirements are as follows:
\begin{itemize}
    \item Understanding the extreme physical conditions that govern star formation and 
    quenching in the early Universe, allowing the formation of the massive galaxies 
    observed at high redshift (e.g., Ref.~\citenum{dekel23});
\item Determining the dark matter (DM) content in high-z galaxies since DM
should drive the assembly of the baryonic matter in the standard model
of galaxy formation (e.g., Ref~\citenum{delucia13});
\item Determining the role played by massive black holes in the formation and evolution of galaxies, 
whether they are the seeds for galaxy formation and powerful extinction mechanisms of star formation
(e.g., Ref~\citenum{finkelstein22});
\item Searching for and, hopefully, detecting the elusive PopIII of primordial stars, necessary 
to address the shortage of elements heavier than Li after the Big Bang and explaining the presence 
of metals in the older stars belonging to PopII (e.g., Ref~\citenum{vanzella23}). 
\end{itemize}
\begin{figure}
\begin{center}
\includegraphics[height=5.5cm]{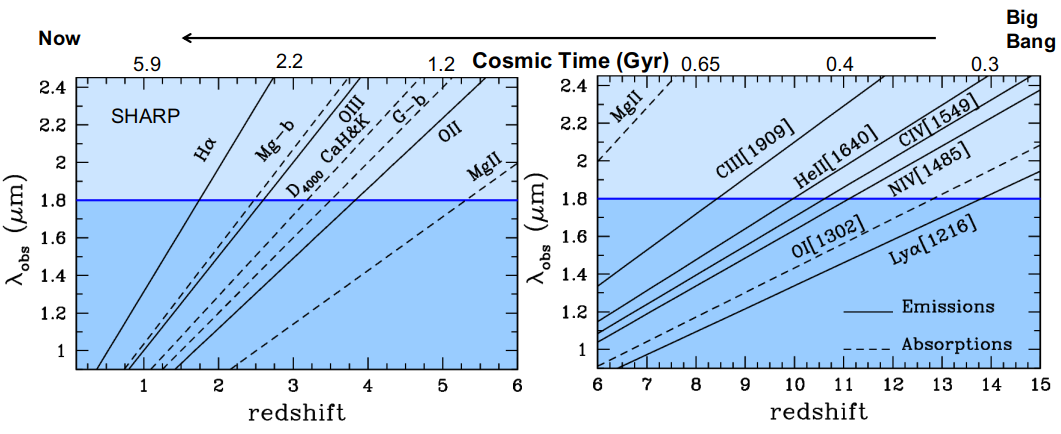}
\end{center}
\caption 
{ \label{fig:lambda}
- The observed wavelength $\lambda_{obs}$=$\lambda_{em}\times(1+redshift)$ of the main nebular 
emission (black solid lines) and absorption (dashed lines)
atomic lines are shown as a function of the redshift of the galaxy. The blue horizontal 
line marks the wavelength limit at $\lambda$=1.8 $\mu$ of the next generation spectrograph 
at ELT (MOSAIC and ANDES) and at VLT (MOONS).} 
\end{figure}
Understanding the physics driving these phenomena needs tracking galaxy properties 
from their early stages to the present. 
These properties, encompassing gas and stellar kinematics, chemical abundances, star 
formation rate and stellar population age, are encoded in the atomic emission and 
absorption lines. 
Most of them requires observations in the near-IR when observed in high-redshift galaxies. 
Specifically, wavelengths longer than $\lambda$$>$1.8 $\mu$m for galaxies at $z$$>$2.5 
(see Fig. \ref{fig:lambda}), coupled with suited spectral resolution (R=$\lambda/{\Delta\lambda}>1000$) 
are requested.
Therefore, the first requirement is to reach the near-IR limit where sky transmission is 
still high and sky emission can be still efficiently removed, i.e. $\lambda_{lim}\simeq$2.45 $\mu$m.

\begin{figure}
\begin{center}
\begin{tabular}{c}
\includegraphics[height=5.5cm]{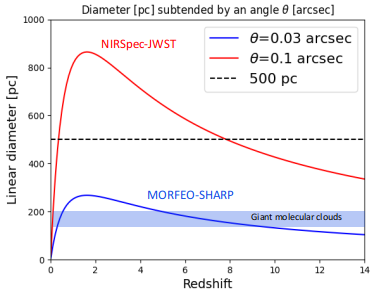}
\end{tabular}
\end{center}
\caption 
{ \label{fig:linearscale}
- The linear size [pc] subtended by an angle $\theta$=0.031" (blue curve), the SHARP pixel scale, 
and by $\theta$=0.1" (red curve), the pixel scale of NirSpec@JWST are shown as a function of redshift. 
The light-blue stripe represents the size of giant molecular gas clouds, 150-200 pc.
The typical half light radius (500 pc) of compact high-redshift galaxies (dashed line), is also shown.} 
\end{figure}

To tackle the above issues, sizes comparable to those of giant molecular gas clouds 
($\sim$150-200 pc) must be resolved.
In fact, such clouds, containing even more than $10^6$ M$\odot$ of molecular gas, 
are the places where massive star formation and enrichment take place, being also 
the best tracers of the galaxy dynamics. 
Fig. \ref{fig:linearscale} shows that an angular resolution of $\sim$30 mas
(blue curve) is needed to resolve giant gas clouds over the whole cosmic time.
Therefore, AO-supported observations coupled with a pixel scale $\sim$30 mass/pix  
are required.

\begin{table}
\caption{- SHARP main requirements resulting from the scientific drivers.} 
\label{tab:requirements}
\begin{center}       
\begin{tabular}{|l|c|} 
\hline
\rule[-1ex]{0pt}{3.5ex}  Wavelength limit & 2.45 $\mu$m  \\
\hline\hline
\rule[-1ex]{0pt}{3.5ex}  Spectral range (simultaneous) & 0.95-2.45 $\mu$m  \\
\hline\hline
\rule[-1ex]{0pt}{3.5ex}  Angular resolution & $\sim$30 mas \\
\hline
\rule[-1ex]{0pt}{3.5ex}  AO-corrected FoV & $\sim$1x1 arcmin  \\
\hline
\rule[-1ex]{0pt}{3.5ex}  Observing modes & MOS + multi-IFU   \\
\hline
\rule[-1ex]{0pt}{3.5ex}  Spectral resolution & R$>$1000 \\
\hline 
\end{tabular}
\end{center}
\end{table}

These measurements should be carried out for many galaxies simultaneously, allowing
us to study protoclusters, over-densities and the multiple clumps seen by JWST in the early Universe.
Therefore, multiplexing capabilities coupled with a large area uniformly corrected for atmospheric
turbulence are required.
These are provided by MORFEO-SHARP. 
Table \ref{tab:requirements} summarizes the main requirements of SHARP.

\section{SHARP}
\begin{figure}
\begin{center}
\begin{tabular}{c}
\includegraphics[height=5.5cm]{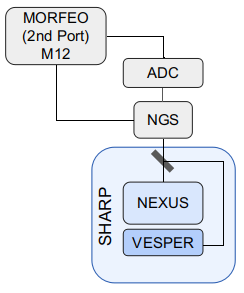}
\end{tabular}
\end{center}
\caption 
{ \label{fig:morfeosharp}
- Components and optical path from the 2nd port of MORFEO to SHARP. 
SHARP is fed by an Atmospheric Dispersion Corrector (ADC) and a 
Natural Guide Stars (NGS) unit.} 
\end{figure}
Figure \ref{fig:morfeosharp} shows the components and the optical path from the 2nd port
of MORFEO to SHARP.
SHARP is fed by an Atmospheric Dispersion Corrector (ADC) and a Natural Guide Stars (NGS) unit. 
Describing the ADC and NGS is beyond the scope of this article. 
Briefly, the light coming from MORFEO will pass through the ADC which allows SHARP to collect 
spectra simultaneously over the whole wavelength range at any zenith distance and rotation angle. 
The NGS wavefront sensors, monitoring up to three stars, pass the information to MORFEO which uses 
also six laser guide stars (LGS) to remove the effects of the atmospheric turbulence over a 
field (at the 2nd port) of diameter 160", encompassing the field of view of SHARP spectrographs. 
The NGS unit will be the same as the one feeding MOSAIC, with the exception of the SCAO
module not used by SHARP.

\begin{figure}
\begin{center}
\begin{tabular}{c}
\includegraphics[height=6.0cm]{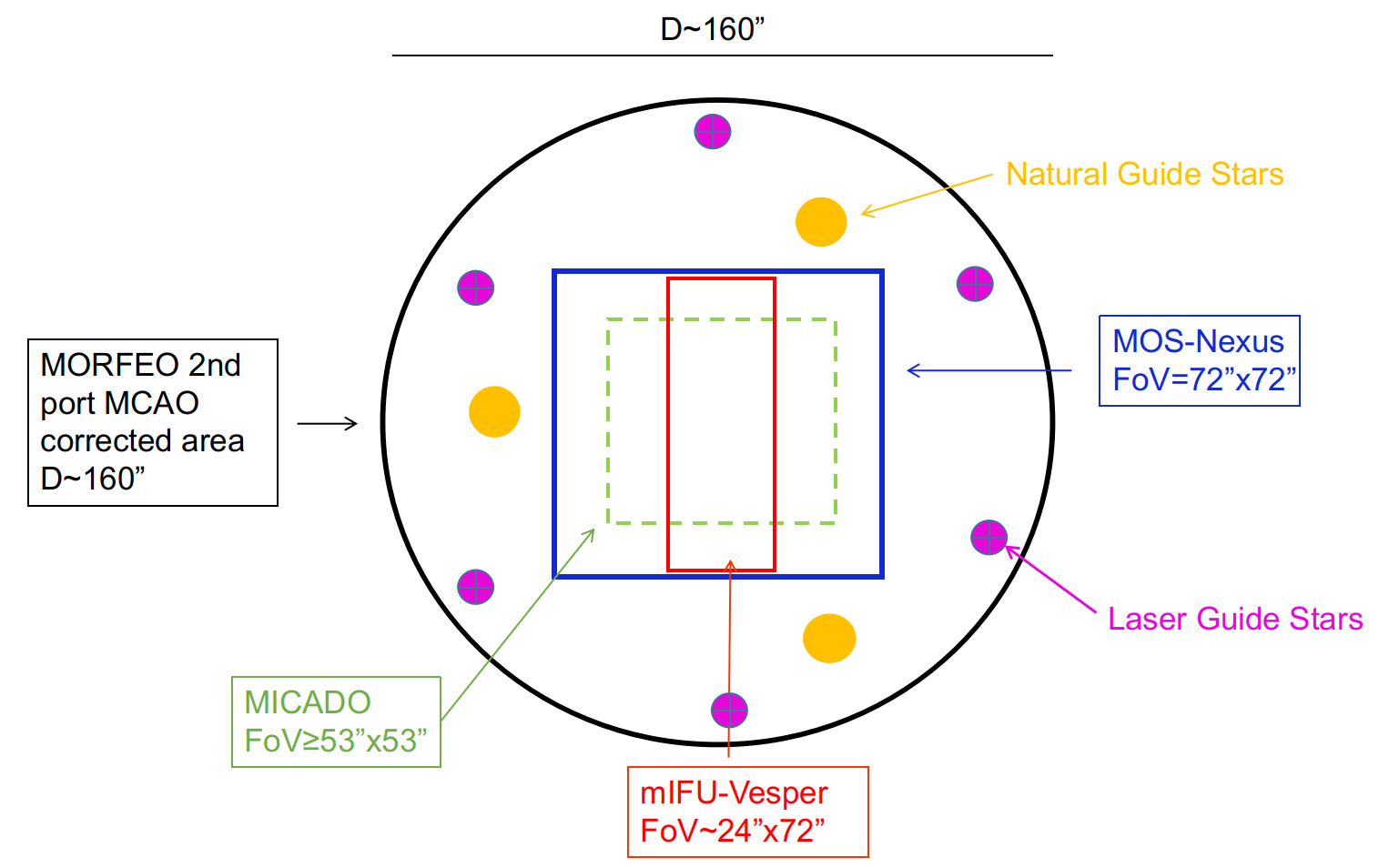}
\end{tabular}
\end{center}
\caption 
{ \label{fig:fov}
- The Field of View (FoV) of NEXUS (blue square, 1.2'$\times$1.2') and the area probed by the 
IFSs of VESPER (red rectangle, $\sim$24"$\times$70") are shown on the MORFEO 2nd port 
corrected area (black circle, diameter D$\sim$160"). 
For comparison, the FoV of MICADO (green dashed square) is also shown.
The purple filled circles represent the wavefront sensors for the 6 Laser Guide Stars used 
by MORFEO, while the yellow circles are those for the 3 Natural Guide Stars.} 
\end{figure}

SHARP consists of two main units: NEXUS, a slitlets Multi-Object Spectrograph (MOS), and VESPER, 
a multi-Integral Field Unit (mIFU).
Given the operative wavelength range SHARP is cryogenic.
Just below the entrance window of SHARP there is the unit selector which
switches the light between NEXUS and VESPER. 
There are no aspheric surfaces in SHARP.

\subsection{NEXUS}
\label{sec:nexus}
NEXUS, the MOS, operates over an AO corrected field $\sim$1.2’$\times$1.2’ (35 mas/pix; see Fig. \ref{fig:fov}).
It is fed by a Configurable Slit System (CSS hereafter) switching among mechanical slit masks able 
of deploying up to 30 slits of $\sim$2.4” length. 
The spectroscopic resolutions for extended sources, i.e., larger than the diffraction-limited Point 
Spread Function (PSF) of ELT ($\sim$12 mas at 2.2 $\mu$m), are R$\sim$6000, $\sim$2000 and $\sim$300
for a reference slit width of 0.2". 
For point sources the resolution is R$\sim$17000. 
The whole wavelength range (0.95-2.45 $\mu$m) is simultaneously covered.
\begin{figure}
\begin{center}
\begin{tabular}{c}
\includegraphics[height=7.0cm]{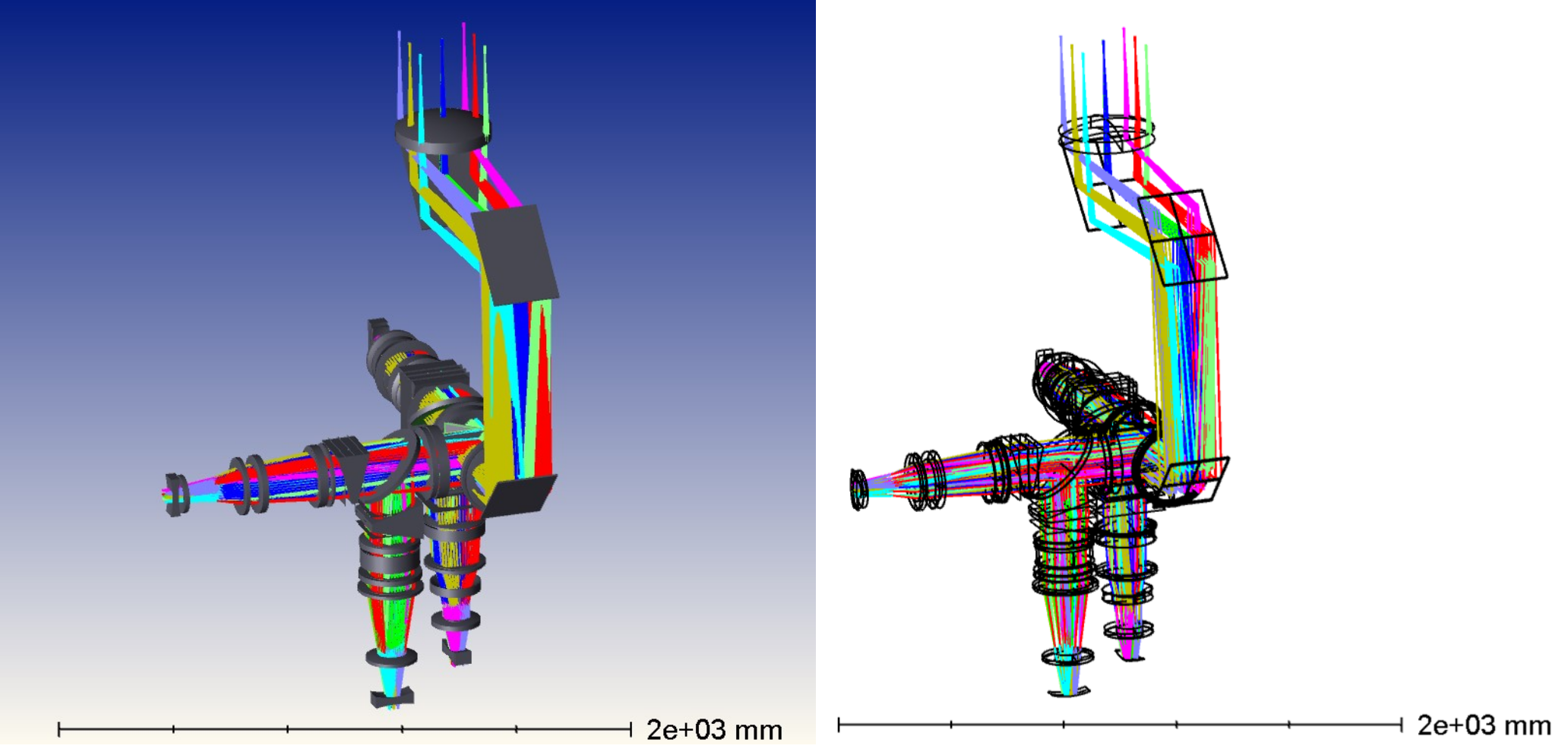}
\end{tabular}
\end{center}
\caption 
{ \label{fig:nexus}
- NEXUS optical design from the focal plane of MORFEO to the four detectors. 
The Configurable Slit System (CSS), not shown in the figure,
is placed onto the focal plane. 
Three folding mirrors provide a compact design, while three dichroics split
the beam into four wavelength ranges feeding 4 cameras.
The whole wavelength range 0.95-2.45 $\mu$m is thus simultaneously covered.} 
\end{figure}

Fig. \ref{fig:nexus} shows the optical design of NEXUS from the focal plane of MORFEO
to the four cameras.
Onto the focal plane of MORFEO there is the CSS whose description is further on.
The light coming from MORFEO and coming out of the CSS is folded 
by three folding mirrors to compact the design and splitted into four wavelength ranges, 
approximately [0.95-1.15]$\mu$m, [1.15-1.45]$\mu$m, [1.45-1.9]$\mu$m, [1.9-2.45]$\mu$m, 
thanks to three dichroics.
Each range is fed by a dedicated camera optimized for the operating wavelength range.
Each camera includes a grism wheel supporting three grisms with resolution 
R$=\lambda/{\Delta\lambda}\simeq$300, 2000 and 6000 for a reference slit with of 0.2".
A cold stop pupil to limit the thermal background feeds the two cameras 
operating at $\lambda>$1.45 $\mu$m.

\begin{figure}
\begin{center}
\begin{tabular}{c}
\includegraphics[height=5.5cm]{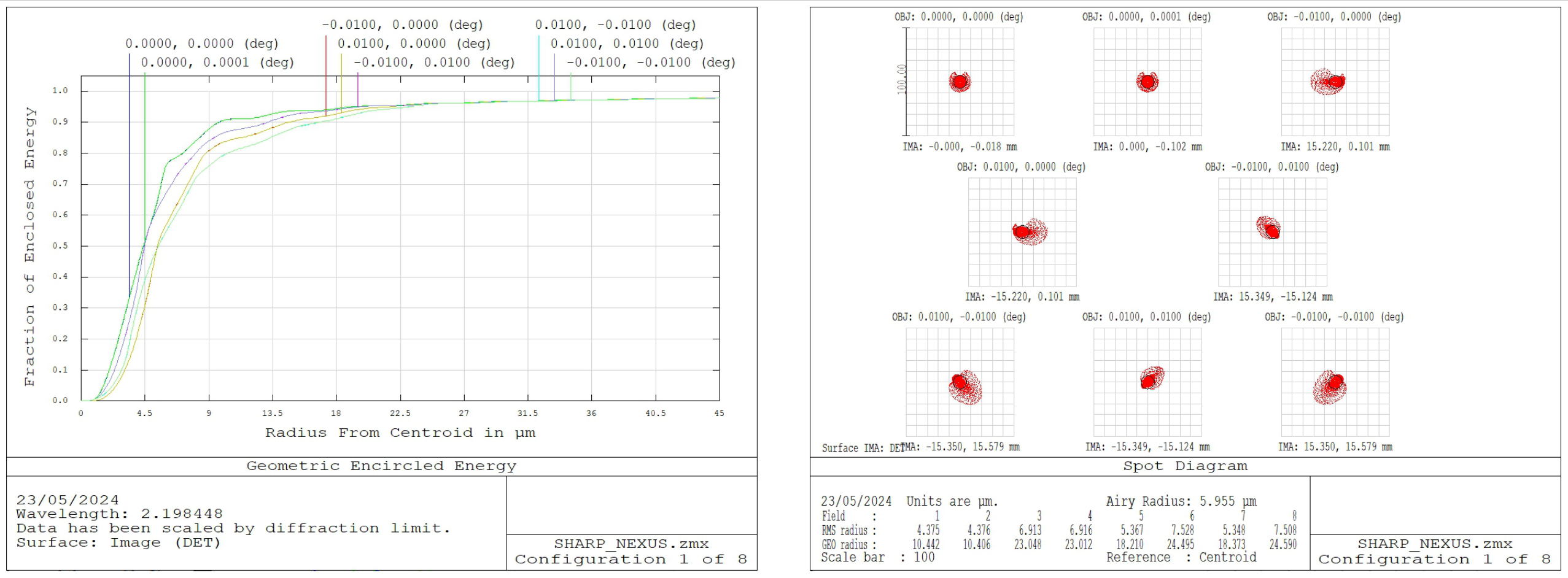}
\end{tabular}
\end{center}
\caption 
{ \label{fig:nexus_EE}
- Left: Fraction of Encircled Energy as a function of radial 
distance for NEXUS at wavelength $\lambda$=2.19 $\mu$m.
The data are scaled for the diffraction limit of ELT.
The curves refer to 5 different positions on the detector.
More than 90\% of the flux fall within one NEXUS pixel (15 $\mu$m, $\sim$35 mas). 
- Right: Spot diagram for the five different positions (central and four offsets)
on the field, as shown in the legend.} 
\end{figure}

 In Fig. \ref{fig:nexus_EE} (left) it is shown the Encircled Energy (EE) distribution as 
 a function of radial distance at wavelength 2.19 $\mu$m (fourth channel) for the central
 position and four positions offset by $\pm$36 arcsec.
 The data are scaled by the diffraction limit of ELT.
 The Figure shows that $>$90\% of the flux fall within one NEXUS pixel 
 (15 $\mu$m, $\sim$35 mas). 

In the right panel of Fig. \ref{fig:nexus_EE} it is shown the spot diagram 
for the different positions on the field obtained.
The mean Airy radius is $\sim$5.9 $\mu$m, almost 1/3 of the pixel size.

\subsubsection{The Configurable Slit System (CSS) of NEXUS}
The CSS of NEXUS is conceptually similar to the one described in Ref.~\citenum{henein04} feeding MOSFIRE at Keck telescope\cite{mclean_mosfire_2012}
and EMIR at GTC telescope\cite{garzon_emir_2022}.
A slit is formed at the position of the selected object by translating two
opposite bars toward each other on the focal plane of NEXUS.
The two bars block the light coming from outside the slit formed by the bars themselves.
Both slit position and slit width are adjustable and are controlled with a micrometric precision \cite{henein04}.
The minimum length of the configurable slit is $\sim$2.4" (the size of the bar being 7.5mm),
according to the recommendation of the SHARP Science Team, this limits the maximum number 
of slits to 30.
Longer slits can be obtained aligning two or more slits, resulting in a slit length multiple of 2.4".
 
\begin{figure}
\begin{center}
\begin{tabular}{c}
\includegraphics[height=5.5cm]{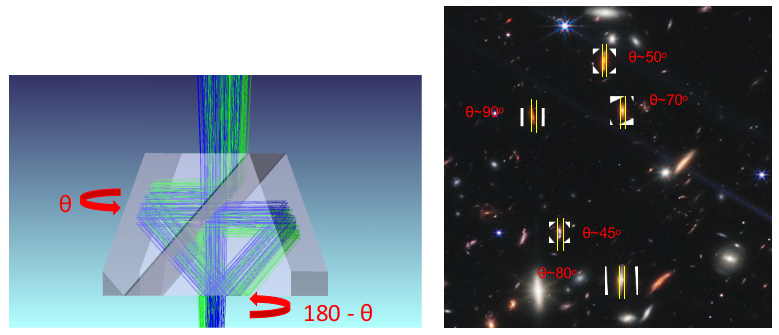}
\end{tabular}
\end{center}
\caption 
{ \label{fig:inversion}
- Left: Optical design of the Pechan inversion prism. 
This prism is composed of two paired prisms.
The light entering from above undergoes five reflections and comes
out from below inverted. 
Rotating the prism by an angle $\theta$ the coming out image is rotated
by a similar angle. Right: Example of an astrophysical application of the inversion prism.
All the target galaxies are aligned along the slit according to their major axis 
to obtain for each of them the rotation curve from which to derive the dynamical mass
of the galaxy (dark matter + baryonic mass).
From the comparison with the mass derived from the stellar light directly related to the 
baryons of the galaxy, the dark matter can be derived addressing one the main scientific
drivers (see Sec. \ref{sect:intro}).} 
\end{figure}

Differently from any other slit system (configurable or not), each slit of NEXUS is 
fed by an inversion prism (see Fig. \ref{fig:inversion}, left panel).
The inversion prism can rotate by an arbitrary angle $\theta$ selected by the user
a square field of $\sim$2.4"$\times$2.4" aligning the target object to the slit,
according to the user's needs.
An example of its application is shown in the right panel of Fig. \ref{fig:inversion}, 
where a number of selected  galaxies are all rotated to align their major axis to the slit.
This allow us to derive for all of them their rotation curve from which
the total mass of each galaxy can be derived.
The comparison with the mass derived from the stellar emission related to the baryonic
mass allow us to derive the dark matter content for all the galaxies.
This makes unique NEXUS and SHARP.

\subsection{VESPER}
\label{sec:vesper}
\begin{figure}
\begin{center}
\begin{tabular}{c}
\includegraphics[height=10.cm]{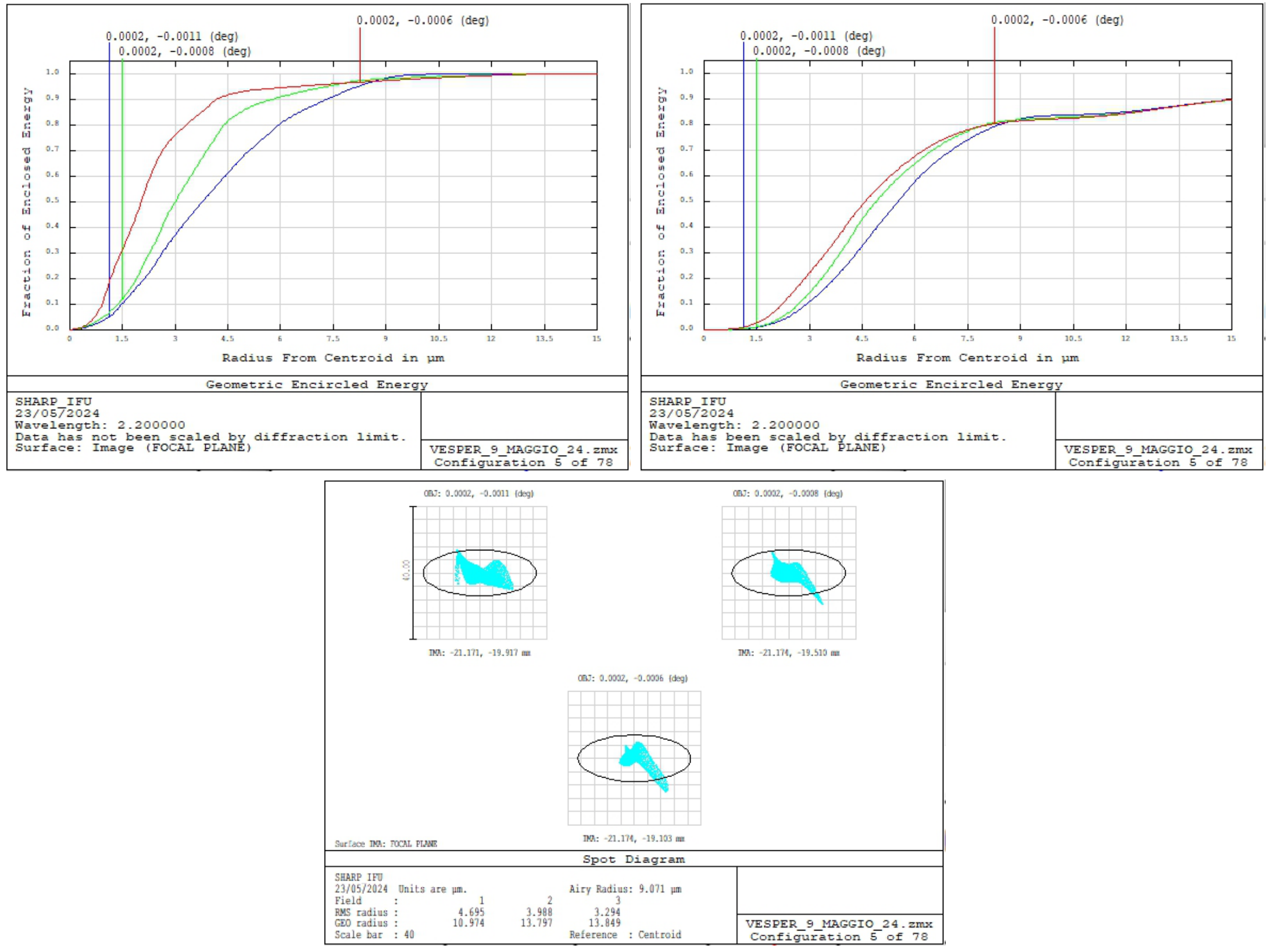}
\end{tabular}
\end{center}
\caption 
{ \label{fig:vesper_EE}
- Upper panel: Left - Fraction of Encircled Energy as a function of radial distance for an
ideal point source not scaled for the diffraction limit of ELT seen by VESPER.
The three curves are for three different positions on one of the 288 mirrors of the
slicer, at the center (red) and at two corners (blue and green).
Right - Same as left panel but scaled for the diffraction limit of ELT. 
Lower panel: Spot diagram measured for the three different positions on the
mirror. 
} 
\end{figure}

VESPER, the multi-IFU, is an image slicer covering simultaneously the wavelength range 1.2-2.4 $\mu$m.
The spectral resolution for extended sources is R$\sim$3000 (R$\sim$4000 for point source).
VESPER is a modular system composed of channels each one comprising 6 probes called Integral 
Field Selectors (IFS hereafter). 
In the current configuration VESPER is composed of two channels summing up to 12 IFSs.
Each IFS has a field of view of $\sim$1.7”$\times$1.5”.
The spaxel scale is 31 mas which maximizes also the SNR 
on the spaxel at $\lambda$=2.2 $\mu$m, according to the performance (encircled energy distribution) of MORFEO.
In the upper panel of Fig. \ref{fig:vesper_EE} it is shown the encircled energy as a function of
the radial distance for thee different positions on a mirror of the slicer for a point source,
for the ideal case of 
no diffraction limit (left panel) and considering the diffraction limit of ELT (right panel).
The comparison between the two provides a direct measure of the quality of the VESPER optical design. It can be seen that the image quality is dominated by the optical systems that precede SHARP.

In the lower panel of Fig. \ref{fig:vesper_EE} it is shown the spot diagram for the
three different positions 
considering the diffraction limit of ELT.

The IFSs are disposed in a cartesian $xy$ grid, separated by 0.33" along the x direction.
These gaps between IFSs are masked on the focal plane to cancel out the contribution 
to the background from scattered light.
The IFSs can be deployed along $\sim$70” in $y$ direction to probe an AO corrected 
area of $\sim$24”x70” 
(see Fig. \ref{fig:fov}).
The equivalent area of the 12 IFS is 1.7”$\times$1.5”$\times$12$\simeq$31 arcsec$^{2}$, 
summing up to 8 detectors 4k$\times$4k.
Figure \ref{fig:ifs}a shows one VESPER channel composed of 6 IFSs.
IFSs can be arranged arbitrarily along the $y$ direction to form $n$ different
fields with FoV=(1.7”$\times$1.5”)*$k$, where $k$ is the number of contiguous IFSs displaced at
the same $y$. 

\subsubsection{The Integral Field Selector (IFS) system}
\begin{figure}
\begin{center}
\begin{tabular}{c}
\includegraphics[height=7.0cm]{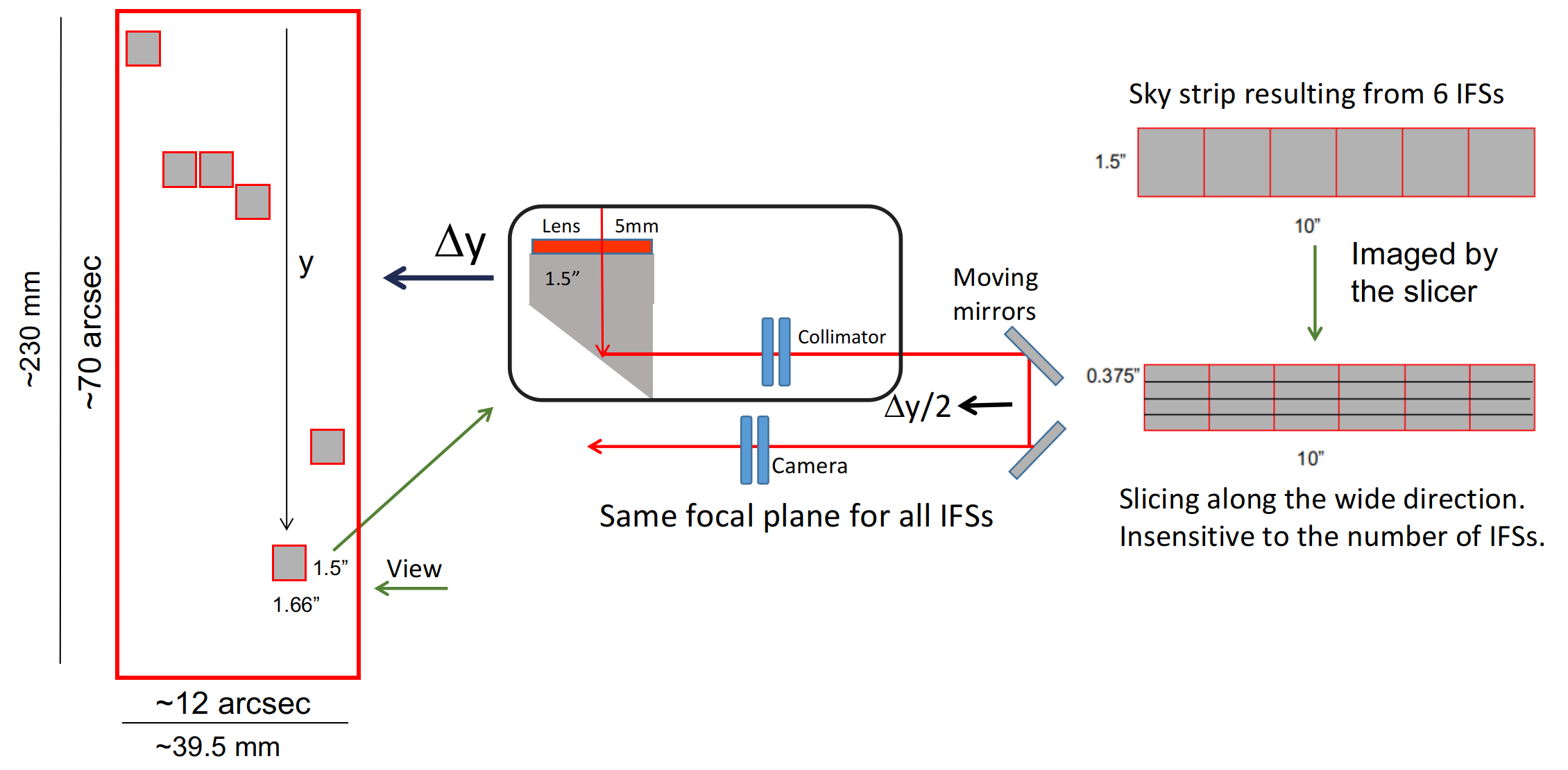}
\end{tabular}
\end{center}
\caption 
{ \label{fig:ifs}
- Schematic view of the Integral Field Selector system feeding one channel of VESPER.
Left - The 6 IFSs are aligned along the $x$ direction and separated by 0.33" each other.
The small gaps between the IFSs are masked on the focal plane to cancel out their contribution 
to the background due to scattered light.
The IFSs can be moved along the $y$ direction to sample an area of $\approx$12"$\times$70".
Center - Each IFS is composed of a prism having the upper surface with power and a collimator. 
They rigidly move by $\Delta y$. 
Their movement is compensated by the movement ($\Delta y/2$) of the two 45-degree 
moving mirrors to keep constant the optical path between collimator and camera for all the IFSs.
Right - The image formed onto the VESPER focal plane by the 6 IFSs is a strip as 
if they were contiguous. 
Onto the VESPER focal plane there is the slicer that sample the
image in slices of 0.375"$\times$10".} 
\end{figure}
The light coming from MORFEO is deviated toward the IFS system by the selector unit.
Figure \ref{fig:ifs} (left) shows the 6 IFSs feeding one channel of VESPER covering half of 
the total area, i.e. $\sim$12”x70”.
Each IFS is composed of a prism and a collimator that rigidly move by $\Delta y$ 
along the $y$ direction (Fig. \ref{fig:ifs} center).
Two 45-degree movable mirrors bend the beam and move by $\Delta y/2$ each to keep the optical 
path between the collimator and camera constant, so that all IFSs share the same focal plane.
On the focal plane the 6 IFSs form an image 1.5"$\times$10" as if they were all 
contiguous (Fig. \ref{fig:ifs} right).
\begin{figure}
\begin{center}
\begin{tabular}{c}
\includegraphics[height=8.0cm, angle=-90]{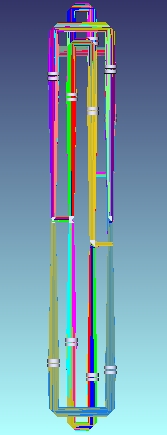}
\end{tabular}
\end{center}
\caption 
{\label{fig:optifs}
- Optical design of the Integral Field Selector system of one channel of VESPER. 
The 6 IFSs move along parallel planes to avoid vignetting.} 
\end{figure}
In Fig. \ref{fig:optifs} the optical design of the IFS system feeding one channel of VESPER (6 IFSs) is shown.
The 6 IFSs move along parallel planes alternatively offset up and down by about 25mm.
Furthermore, the prisms are oriented two by two to send the light in opposite directions.
This design avoid interference and vignetting among the components.

\subsubsection{The slicer}
\begin{figure}
\begin{center}
\begin{tabular}{c}
\includegraphics[height=6.0cm]{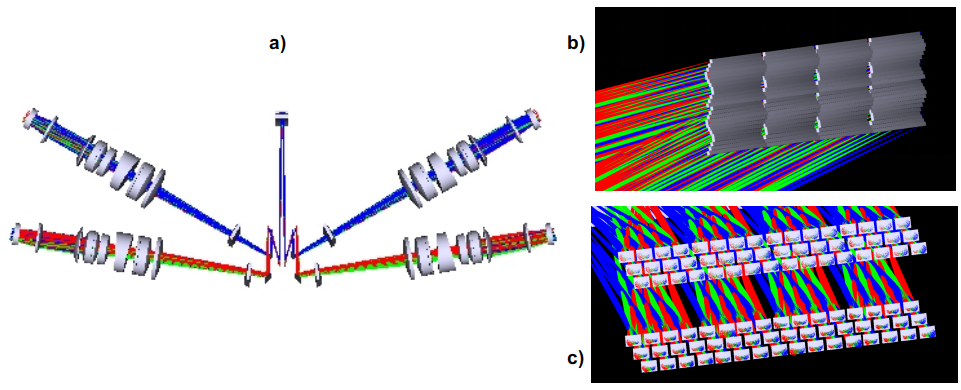}
\end{tabular}
\end{center}
\caption 
{\label{fig:slicer}
- a) Optical design of one channel of VESPER, from the IFS system to the 4 cameras.
Each camera feeds a set of mirrors of the slicer.
b) Slicer onto the focal plane of VESPER. The slicer is composed of four set of 72 mirrors each.
Two set are shown as example.
Each set samples a slice of 0.375"$\times$10" of the image provided by the 6 IFSs.
c) Particular of two set of pupil mirrors bringing the light to the corresponding cameras.
} 
\end{figure}

The light coming from the IFS system of VESPER is focused on the slicer, 
conceptually similar to the slicer of MUSE at VLT (Ref.~\citenum{henault04}).
The slicer of VESPER is composed of four sets of 72 mirrors (the slicer of MUSE is made of 
sets of 48 mirrors).
The mirror sizes are 1.0mm$\times$22 mm.
Figure \ref{fig:slicer}b shows the design of two sets of mirrors.
Each set of 72 mirrors samples a slice 0.375"$\times$10" of the image along the wide direction.
There is a corresponding set of pupil mirrors (Fig. \ref{fig:slicer}c) for each set of mirrors 
that converges the light toward the camera where there is the spectrograph.
The signal from each spaxel (31 mas) is fed into the spectrograph that generates a spectrum for each one of them. 
The spectral resolution is R$\sim$3000 for extended sources and R$\sim$4000 for point sources. 
There is a camera feeding each set of 72 mirrors, therefore 4 cameras (see Fig. \ref{fig:slicer}a)
for one channel (6 IFSs), summing up to 8 cameras for the two channels of VESPER.

Table \ref{tab:properties} summarizes the main properties of NEXUS and VESPER.
\begin{table}
\caption{- Properties of the Multi-Object Spectrograph NEXUS and the Multi-Integral Field Unit VESPER
of SHARP.} 
\label{tab:properties}
\begin{center}       
\begin{tabular}{|l|c|c|} 
\hline
 &  {\bf NEXUS}&  {\bf VESPER}\\
\hline\hline
Spectral range (simultaneous) & 0.95-2.45 $\mu$m  & 1.2-2.4 $\mu$m \\
\hline\hline
FoV/Area probed & 1.2'$\times$1.2'  & 24"$\times$70"\\
\hline
Multiplexing & up to 30 slits (2.4" slit length)& 12 IFSs (1.7"$\times$1.5" each)\\
\hline
Pixel scale & 35 mas  & 31 mas \\
\hline
Spectral resolution (ext. source) & 300, 2000, 6000 & 3000 \\
\hline 
Spectral resolution (point source) & $\sim$17000 & $\sim$4000 \\
\hline 
\end{tabular}
\end{center}
\end{table}


\section{Conclusions}
In this paper, we presented the scientific motivations of the SHARP near-IR multimode spectrograph
and the consequent main requirements that guided the optical design and some opto-mechanical
choices.
SHARP is designed for the 2nd port of MORFEO to exploit the capabilities of ELT, i.e. 
the world's largest collecting area and unmatched angular resolution.
We have reached an advanced and stable optical design whose opto-mechanical design is ongoing.
Most of the key optical performances have already been verified through qualification models.

The final aim of the project is to propose SHARP to ESO's next call for new instrumentation. 
In this view, we are spreading SHARP to the scientific and technological community to generate 
interest and find partners. 
Our plan is to assemble an international team and consortium and to carry out a complete feasibility 
study of the instrument by mid 2026. 
More information, contacts and tools can be found at the website of the
SHARP project http://sharp.brera.inaf.it.

MORFEO-SHARP exceeds the observational limits fixed by NIRSpec@JWST allowing us to explore the 
new paths that JWST is opening. 
SHARP can take up the baton left by JWST when its mission ends.

\subsection*{Acknowledgments}
The SHARP team acknowledges support by grant "Bando INAF Ricerca Fondamentale 2022", Techno-Grant SHARP - 1.05.12.02.01. PS would like to thank V. Lynd for the useful discussions.



\bibliography{sharp_bib}   
\bibliographystyle{spiejour}   


\vspace{2ex}\noindent\textbf{Paolo Saracco} is Prime Researcher at INAF - Osservatorio Astronomico di Brera in Milan, Italy. 
He received his MS degrees in 
Astronomy from the University of Bologna in 1992 and his PhD degree in 
Astrophysics from the University of Milano 1995.  
He is the author of more than 85 refereed journal papers on galaxy evolution. 
His current research interests include instrumentation for the ELT.


\end{spacing}
\end{document}